\documentstyle[11pt]{article}
\normalbaselines
\renewcommand{\baselinestretch}{1.2}
\renewcommand{\theequation}{\thesection.\arabic{equation}}
\newtheorem{thm}{Theorem}[section]
\newcommand{\be}{\begin{equation}}
\newcommand{\ee}{\end{equation}}  
\newcommand{\pri}{\prime}  
\newcommand{\bi}[1]{
\renewcommand{\arraystretch}{0.4} (\!\! \ba{c} 
\scriptscriptstyle {#1} \\ \scriptscriptstyle  2\ea\!\!)
\renewcommand{\arraystretch}{1} }
\newcommand{\re}{\mbox{Re$\;$}}
\newcommand{\im}{\mbox{Im$\;$}}
\newcommand{\mg}{\rm}
\renewcommand{\em}{\it}
\newcommand{\D}{{\cal D}_{q}}
\newcommand{\E}{{\cal E}_q}
\newcommand{\Q}{{\cal Q}}
\renewcommand{\a}{\alpha}
\renewcommand{\b}{\beta}
\renewcommand{\l}{\lambda}
\newcommand{\f}{\phi}
\newcommand{\plus}{\overset{\circ}{+}}
\renewcommand{\t}{\theta}
\renewcommand{\r}{\rho}
\newcommand{\g}{\gamma}
\newcommand{\dis}{\vspace{-0.5\abovedisplayskip}}
\newcommand{\bea}{\begin{eqnarray}}
\newcommand{\eea}{\end{eqnarray}}
\newcommand{\ac}{ Al-Salam-Chihara }
\newcommand{\acp}{ Al-Salam-Chihara polynomials }
\newcommand{\Sum}{\sum_{n=0}^\infty}
\newcommand{\aw}{ Askey-Wilson  }
\newcommand{\gauss}[2]{\genfrac{[}{]}{0pt}{}{#1}{#2}_q} 
\newcommand{\awp}{ Askey-Wilson polynomials }
\newcommand{\awo}{ Askey-Wilson operator }

\begin{document}
\bibliographystyle{unsrt}

\newtheorem{theorem}{ Theorem}
\newtheorem{lemma}{Lemma}
\newtheorem{proposition}{Proposition}

\def\bea*{\begin{eqnarray*}}
\def\eea*{\end{eqnarray*}}
\def\ba{\begin{array}}
\def\ea{\end{array}}
\count1=1
\def\be{\ifnum \count1=0 $$ \else \begin{equation}\fi}
\def\ee{\ifnum\count1=0 $$ \else \end{equation}\fi}
\def\ele(#1){\ifnum\count1=0 \eqno({\bf #1}) $$ \else \label{#1}\end{equation}\fi}
\def\req(#1){\ifnum\count1=0 {\bf #1}\else \ref{#1}\fi}
\def\bea(#1){\ifnum \count1=0   $$ \begin{array}{#1}
\else \begin{equation} \begin{array}{#1} \fi}
\def\eea{\ifnum \count1=0 \end{array} $$
\else  \end{array}\end{equation}\fi}
\def\elea(#1){\ifnum \count1=0 \end{array}\label{#1}\eqno({\bf #1}) $$
\else\end{array}\label{#1}\end{equation}\fi}
\def\cit(#1){
\ifnum\count1=0 {\bf #1} \cite{#1} \else 
\cite{#1}\fi}
\def\bibit(#1){\ifnum\count1=0 \bibitem{#1} [#1    ] \else \bibitem{#1}\fi}
\def\ds{\displaystyle}
\def\hb{\hfill\break}
\def\comment#1{\hb {***** {\em #1} *****}\hb }
\newcommand{\TZ}{\hbox{\bf T}}
\newcommand{\MZ}{\hbox{\bf M}}
\newcommand{\ZZ}{\hbox{\bf Z}}
\newcommand{\NZ}{\hbox{\bf N}}
\newcommand{\RZ}{\hbox{\bf R}}
\newcommand{\CZ}{\,\hbox{\bf C}}
\newcommand{\PZ}{\hbox{\bf P}}
\newcommand{\QZ}{\hbox{\bf Q}}
\newcommand{\HZ}{\hbox{\bf H}}
\newcommand{\EZ}{\hbox{\bf E}}
\newcommand{\GZ}{\,\hbox{\bf G}}

\font\germ=eufm10
\def\goth#1{\hbox{\germ #1}}
\newtheorem{pf}{Proof}
\renewcommand{\thepf}{}
\vbox{\vspace{38mm}}

\begin{center}
{\LARGE \bf Bethe Ansatz Equations of XXZ Model
and \\ [2mm] q-Sturm-Liouville Problems}\\[15mm]

Mourad E. H. Ismail \\ {\it Department of
Mathematics, University of Central Florida \\
Orlando, Florida USA 32816 \\ 
(e-mail:ismail@math.usf.edu)}
\\[2mm]
Shao-shiung Lin
\\ {\it Department of Mathematics,  
Taiwan University \\ Taipei, Taiwan \\
(email: lin@math.ntu.edu.tw)}\\[2 mm]
Shi-shyr Roan  \\{\it
Institute of Mathematics, Academia Sinica \\  Taipei, Taiwan \\ (e-mail:
maroan@gate.sinica.edu.tw)} 
\\ [8mm]
\end{center}

\begin{abstract} 
In this article we have discovered 
a close relationship between the
(algebraic) Bethe Ansatz equations of the spin $s$ XXZ model
of a finite size and the $q$-Sturm-Liouville
problem. We have demonstrated that solutions of
the Bethe Ansatz equations give rise to  the
polynomial solutions of a second order
$q$-difference equation in terms of Askey-Wilson
operator. The more general form of 
Bethe Ansatz equations and the mathematical problems
relevant to the physics of XXZ model are
discussed. Furthermore, the similar correspondence
between Bethe Ansatz equations of XXX model and
the Sturm-Liouville type difference equation in
terms of Wilson operator has also been found. 
\end{abstract}
\par \vspace{3mm} \noindent
  \par \noindent
1999 PACS: 02.30Gp, 05.50.+q, 75.10.Jm.  \par \noindent

\section{Introduction}

Among the many one-dimensional integrable quantum
spin chains the XXZ spin chain is an  important and renowned  one. It 
corresponds to the 6-vertex model in the 2-dimensional solvable
statistical mechanics. In particular, the 
XXZ Hamiltonian of spin $\frac{1}{2}$ of a finite
size $2N$ has been the subject of extensive
studies for a long time in the physics 
community, and in  recent years it has been investigated by
mathematicians in the areas of mathematical
physics and quantum algebras. The one-dimensional 
$U_q(sl_2(\CZ))$-invariant XXZ model of
spin $\frac{1}{2}$ of a size $2N$ with the
open (Dirichlet) boundary condition  is
described by the following
Hamiltonian \cite{S}, \cite{KS},  
\be
 H_{\rm XXZ}^{(o)} = - \sum_{j=1}^{2N-1} (
\sigma_j^1 \sigma_{j+1}^1 +
\sigma_j^2 \sigma_{j+1}^2 + \triangle 
\sigma_j^3 \sigma_{j+1}^3 ) -
\frac{q-q^{-1}}{2}(\sigma_1^3-
\sigma_{2N}^3),
\
\
\ \triangle := \frac{q+q^{-1}}{2}, 
\ele(XXZ)
where $\sigma^j_n$ are the Pauli matrices acting
on the $j^{\it th}$ site:  
$$ \sigma^1 = \left( \begin{array}{cc}
          0 & 1 \\
          1 & 0
         \end{array}   \right), \ \
\sigma^2 = \left( \begin{array}{cc}
          0 & -{\rm i} \\
         {\rm i} & 0
         \end{array}   \right), \ \
\sigma^3 = \left( \begin{array}{cc}
          1 & 0 \\
          0 & -1
         \end{array}   \right) . 
$$

The Hamiltonian $ H_{\rm XXZ}^{(o)}$ defines
a linear endomorphism of  $\stackrel{2N}{\otimes}
\CZ^2$, whose eigenvalue problem has been the main
concern for the physical quantities related to
the system. In the context of quantum inverse
scattering method developed by the Leningrad
school in the early eighties (see, for example 
\cite{F}, \cite{KS8}), the diagonalization problem
of the Hamiltonian has been investigated by means
of solutions of the following algebraic Bethe
Ansatz equations,
\begin{eqnarray*}
\left(\frac{\sin(\lambda_k+ \frac{1}{2} \eta)}{\sin
(\lambda_k-
\frac{1}{2} \eta) } \right)^{2N}=
\prod_{j
\neq k,  j=1}^n \frac{\sin
(\lambda_k+\lambda_j+\eta)
\sin (\lambda_k-\lambda_j+\eta)}{\sin (\lambda_k+\lambda_j-\eta)
\sin (\lambda_k-\lambda_j-\eta)}, \ \ \ 1 \leq k
\leq n.
\end{eqnarray*}
In general, for a given positive half-integer $s$,
the theory also provides a similar XXZ
Hamiltonian of spin
$s$ of a finite size $2N$  with
the 
following form of  Bethe Ansatz
equations associated to the corresponding
Hamiltonian,
\be
\left(\frac{\sin(\lambda_k+s \eta)}{\sin (\lambda_k-
s \eta) }\right)^{2N}=
\prod_{j
\neq k,  j=1}^n \frac{\sin
(\lambda_k+\lambda_j+\eta)
\sin (\lambda_k-\lambda_j+\eta)}{\sin (\lambda_k+\lambda_j-\eta)
\sin (\lambda_k-\lambda_j-\eta)}, \ \ \ 1 \leq k
\leq n.
\ele(BetheXXZ)

For over half a century  the Bethe Ansatz has been a very 
useful tool in obtaining important informations of
the XXZ-model in  physics literature; but yet there is no 
systematic  study of the mathematical content
on the Bethe Ansatz available at this time.  
In this paper, we establish an explicit connection 
between the
relation (\req(BetheXXZ)) and $q$-Sturm-Liouville
problem. In fact, we study the Bethe Ansatz equations in a
more general setting than (\req(BetheXXZ)), namely the system of equations 
\be
\prod_{l=1}^{2N}
\frac{\sin(\lambda_k+s_l \eta)}{\sin (\lambda_k-
s_l \eta) } =
\prod_{j
\neq k,  j=1}^n \frac{\sin
(\lambda_k+\lambda_j+\eta)
\sin (\lambda_k-\lambda_j+\eta)}{\sin (\lambda_k+\lambda_j-\eta)
\sin (\lambda_k-\lambda_j-\eta)}, \quad 1 \leq k
\leq n,
\ele(gBA)
where $s_l$'s are $2N$ complex numbers. The
mathematical problem is to obtain the solution(s)
of the above  system of nonlinear equations. The
solution of (\req(gBA)) will determine the roots
of a polynomial which satisfies a $q$-difference relation which is a  
Sturm-Liouville type equation involving  the
Askey-Wilson operator, as we shall see in \S 4. 
For $N=2$, the system of equations   
(\req(gBA)) is solved by the zeros of the 
Askey-Wilson polynomials. An interesting physical problem is to 
understand  the large $N$ behavior of solutions  
 of the system of equations (\req(gBA)). Such a study 
is  a challenging  problem in
the area of $q$-Sturm-Liouville equations. The
understanding of the solutions of
(\req(gBA)) and their limiting 
distribution will have a profound impact on the physics of statistical mechanics.

As $q \to 1$, it is known
that the XXZ chain of spin $s$ becomes the
$sl_2(\CZ)$-invariant spin $s$ XXX chain of a
finite size $L=2N$ with the periodic condition. 
The antiferromagenetic  spin
$\frac{1}{2}$  XXX  chain of size $L$ with
periodic condition is given by the following 
Heisenberg XXX Hamiltonian:
\be
 H_{\rm XXX} = - J \sum_{j=1}^L (
\sigma_j^1 \sigma_{j+1}^1 +
\sigma_j^2 \sigma_{j+1}^2 +  
\sigma_j^3 \sigma_{j+1}^3  -1 ), \ \ \ J < 0.
\ele(XXX)        
The above XXX spin chain is a 
famous integrable model with many 
applications to solid state physics and
statistical mechanics. It was first proposed 
by Heisenberg in 1928, \cite{He}, then solved by Bethe
in 1931 \cite{Be}. The spectral 
problem of $H_{\rm XXX}$ can be reduced to the solution of 
the following Bethe Ansatz equations,
$$
\left(\frac{\lambda_k +  \frac{\rm i}{2}}{\lambda_k
- \frac{\rm i}{2}}\right)^L = \prod_{j=1, j \neq k}^l
\frac{\lambda_k - \lambda_j + {\rm i}}{\lambda_k
- \lambda_j - {\rm i}}, \ \ \lambda_k \in \CZ, \ \ \ k = 1, \ldots, l.
$$
We shall discuss the above equations subject to 
certain symmetry conditions imposed  on the roots
$\lambda_j$'s from physical
considerations of the ground state of 
Hamiltonian $H_{\rm XXX}$. In this situation, the
Bethe Ansatz equations are closely connected to a
Sturm-Liouville problem involvong a Wilson
operator.  The solution of this problem requires new
developments in the theory of Wilson
operators. Further mathematical study of those difference
equations could enrich our understanding of both the
mathematical and physical contents related to the
Bethe Ansatz equations of XXX model. This will 
be the subject of a future work. 

An important contribution of this paper is to point  out that 
the Bethe Ansatz equations are variations on nineteenth century work
by  Heine, Stieltjes and Hilbert. Heine studied polynomial solutions
to a second order differential equations 
\be
\Pi(x) \frac{d^2 y}{dx^2}  + \Phi(x) 
\frac{dy }{dx} + r(x) y =0 .
\ele(HS)
In (\req(HS)),  $\Pi$ and $\Phi$ are given polynomials of degrees 
$N$ and $N-1$. Heine proved that given a nonnegative integer  $n$, 
there exists at most ${N+n-2 \choose N-2}$ choices of the polynomial 
$r$ in (\req(HS)) such that (\req(HS)) has a polynomial solution. 
Stieltjes continued this research and showed that if we assume that 
$\Pi$ and $\Phi$ have only real and simple zeros and their zeros 
interlace, then there are precisely  ${N+n-2 \choose N-2}$
polynomials 
$r$ which will make (\req(HS)) have a polynomial solution of degree $n$. 
For references and details we refer the interested reader to Szeg\H{o}'s 
book \cite{Sz}, which also treats Hilbert's work on the location of zeros of 
Jacobi polynomials $P_n^{(\a, \b)}(x)$ when the conditions of orthogonality, 
namely ${\rm Re} \ \a > -1$,  Re $\b > -1$, and $\a + \b$ real, 
are not necessarily assumed. 
If we denote the zeros of a polynomial solution of (\req(HS)) by 
$x_1, \dots, x_n$ then when $x = x_k$, for $1 \le k \le n$, equation 
(\req(HS)) becomes 
\be 
 \sum_{1 \le j \le n, j \ne k} \quad 
\frac{1}{x_j-x_k} = \frac{\Phi(x_k)}{2 \Pi(x_k)}. 
\ele(SBA) 
The system of algebraic equations (\req(SBA)) is a system
of generalized  Bethe Ansatz equations. Observe that although the
polynomial $r$ does  not appear explicitly in the system
(\req(SBA)), it is used implicitly to  show the existence of a
polynomial solution to (\req(HS)),  hence the zeros of the
polynomial $y$ solve (\req(SBA)). Conversely if  (\req(SBA)) has a
solution $x_1, \dots, x_n$, then  we set 
$y(x) = \prod_{k=1}^n (x - x_k)$ and observe that 
$\Pi(x) y''+ \Phi(x) y'$ vanishes at the zeros of $y$ 
hence there is a polynomial $r$ of degree (at most)
$N-2$ such that (\req(HS)) holds. This shows that the number 
of different solutions to (\req(SBA)) is the same as the number 
of choices of the polynomial 
$r$ in (\req(HS)). It is exactly this set up that is behind the modern Bethe 
Ansatz equations where second order differential equations are
replaced by second orderequations in the\aw operator (XXZ model) or
the Wilson operator (XXX model).

The paper is organized
as follows.  In Section  2, we indicate the equivalence of
(\req(SBA)) to  second order equations in the\aw operator, which is
a system of equations  that generalize the Bethe Ansatz equations
for the XXZ model.  In Section 3 we provide  intuitive explanations
for closed form polynomial solutions to second order  differential
equations with polynomial coefficients of degrees  2, 1, and zero,
as well as similar equations  where derivatives are replaced by
applications of  
$q$-difference and\aw operators. This explains where the big $q$-Jacobi 
polynomials and the\aw polynomials come from. 

Section 4 is devoted to studying the general second order equation in the\aw 
operator  $\D$ with general polynomial coefficients. We identify the symmetric 
form of such operator equation through an amazing simplification resulting from 
expanding the coefficients in Chebyshev polynomials of the first and second 
kinds. These representations lead in 
Section 5 to the concept of regular singular 
points of the second order operator equation in an\aw operator with polynomial 
coefficients. 

In Section  6, we consider the  case corresponding to
$q = 1$ in 
Section 4. The relationship of the Sturm-Liouville
problem in terms of Wilson operator and the Bethe
Ansatz equations of the XXX model for the ground state
has been found.

 {\bf Convention}.  In this
paper, 
$\ZZ, \RZ, \CZ$ will denote 
the ring of integers, real, complex numbers
respectively, $\NZ = \ZZ_{>0}$, 
$\RZ^*=
\RZ \ \setminus \ \{0\}$, $\CZ^* =\CZ \ \setminus
\ \{0\}$, and ${\rm i} = \sqrt{-1}$.

\setcounter{equation}{0}

\section{q-Sturm-Liouville Problems}

Given a function $f$, we set
$\breve{f}(z) = f(x)$ with 
$$
z= e^{{\rm i}
\theta}, \ \ \ x =\cos \theta = (z+z^{-1})/2.
$$
Indeed $z = x \pm \sqrt{x^2-1}$ with the branch of the square root 
chosen to make $\sqrt{x^2-1} \approx x$, as $x \to \infty$. This makes 
$|z^{-1}| \le 1 \le |z|$, and $=$ holds if and only if $x \in [-1,1]$. 
With this notation we now introduce the following operations on $f(x)$,
\begin{eqnarray}
(\eta_q f) (x) &=& \breve{f} ( q^{1/2} z), \nonumber \\ 
({\cal D}_q f) (x) &=& \frac{(1-x^2)^{-1/2}}{{\rm i}
(q^{1/2}- q^{-1/2})} (\eta_q f -\eta_{q^{-1}}f)(x), \label{Ddef}
\\
({\cal A}_q f) (x) &=& \frac{1}{2}(\eta_q f +
\eta_{q^{-1}}f)(x) . \nonumber
\end{eqnarray} 
The operator ${\cal D}_q$ is called the
Askey-Wilson operator \cite{Ga:Ra}, \cite{Is}. It is important
to observe that both ${\cal D}_q$ and ${\cal A}_q$ are
invariant under $q \to q^{-1}$.  

Recall  that the Chebyshev
polynomials of the first and second kinds,  respectively,  are
\be
T_n(\cos \t) = \cos n\t, \qquad U_n(\cos \t) = 
\frac{\sin(n+1)\t}{\sin \t}.
\ee
The\awo has the properties 
$$
\D T_n(x) = \frac{q^{n/2} -q^{-n/2}}{q^{1/2}- q^{-1/2}}\; U_{n-1}(x), 
\quad {\cal A}_q T_n(x) = \frac{1}{2}\left[q^{n/2}+q^{-n/2}\right]T_n(x). 
$$
Thus $\D$ reduces the degree of a polynomial by 1 while ${\cal A}_q$ 
preserves the degree of a polynomial. 
Furthermore the\awo has the following properties, \cite{Is},
\bea(ll)
{\cal D}_q (fg) &= (\eta_q f)( {\cal D}_qg) +
(\eta_{q^{-1}}g )({\cal D}_qf) \\
&= ({\cal A}_q f)( {\cal D}_qg) +
({\cal A}_q g )({\cal D}_qf) .
\elea(Dfg)
The second line in the above equation follows from the first by interchanging 
$f$ and $g$ in the first line then taking the average of the two answers. 

Thus ${\cal D}_q, {\cal A}_q$ are
operators of the polynomial algebra $\CZ[x]$ with 
${\rm deg}\, {\cal D}_q(f) = {\rm deg}\, f -1$, and 
${\rm deg} \,{\cal A}_q(f) = {\rm deg} \,f$ for $f
\in \CZ[x]$. Furthermore these operators
preserve $\RZ[x]$ for a real $q$.

For convenience, we shall use the
following convention throughout this paper,
\be
q= e^{2 {\rm i} \eta }, \ \  \theta = 2
\lambda, \ \ ({\rm hence} \ x = \cos 2 \lambda ).
\ee
For given   functions
$w(x)$, $p(x)$, $r(x)$, we consider the following
$q$-Sturm-Liouville equation of $f(x)$, 
\be
\frac{1}{w(x)} {\cal D}_q \left(\left(p(x) {\cal D}_q  
\right)f \right)(x) = r(x) f(x).  
\ele(SL)
By (\req(Dfg)), one can rewrite the equation 
(\req(SL)) in the following form,
\be   
\Pi(x) {\cal D}_q^2  f
(x) +\Phi(x)({\cal
A}_q{\cal D}_q f)(x) = r(x) f(x), 
\ele(SLeq)
where the functions $\Pi, \Phi$ are defined by 
\be
\Pi(x)= \frac{1}{w(x)} {\cal A}_q
p(x), \ \ \Phi(x)= \frac{1}{w(x)} {\cal D}_q
p(x) .
\ele(Pi)
The form (\req(SL)) is the symmetric form of (\req(SLeq)), which can 
be seen from the formula of integration by parts in \cite{Br:Ev:Is} which 
will be stated later as (\req(IBP)). 
We shall 
show in \S 4 how to construct $w$ and $p$ from $\Pi$ and $\Phi$. Now 
the relationship 
\begin{eqnarray}
({\cal A}_q{\cal D}_q f)(x) &=& \frac{q\;  e^{{\rm i}\theta} }
{(q-1)(qe^{2{\rm i}\theta}-1)(e^{2{\rm
i}\theta}-q)}
 \nonumber \\
 &{}& \qquad \times   \{(e^{2{\rm
i}\theta}-q)\eta_{q^2}- (qe^{2{\rm
i}\theta}-1)\eta_{q^{-2}} + 
(q-1)(e^{2{\rm
i}\theta}+1)  \} f (x);   \nonumber \\
{\cal D}_q^2 f(x) &=&
\frac{2q^{3/2}\; e^{{\rm i}\theta} }{{\rm
i}(1-q)^2
\sin \theta (q e^{2{\rm i}\theta}-1)(e^{2{\rm
i}\theta}-q)} 
 \nonumber \\
 &{}& \qquad \times 
\{ (e^{2{\rm i}\theta} -q)
\eta_{q^2} +(qe^{2{\rm i}\theta}
-1)\eta_{q^{-2}} -
(q+1)(e^{2{\rm i}\theta} -1)\}f(x), \nonumber
\end{eqnarray}
shows that a root 
$x_0= \cos 2\lambda_0 $ of the polynomial $f(x)$,
$f(x_0)=0$, necessarily satisfies the following
equation,
\begin{eqnarray}
&{}& \{(e^{4{\rm i}\lambda_0}
-q) (\Pi(x_0)
-\Phi(x_0) \sin \eta \sin 2\lambda_0 
)\eta_{q^2}\} f (x_0)    \nonumber \\
&{}&  \qquad + \{(qe^{4{\rm i}\lambda_0}
-1) ( 
\Pi(x_0)
+ \Phi(x_0) \sin \eta \sin 2\lambda_0  )
\eta_{q^{-2}} \}   f (x_0)  = 0,  \nonumber
\end{eqnarray} 
or equivalently,
\be
\left(\frac{\eta_{q^2}f}{\eta_{q^{-2}}f}\right)(x_0) = 
\frac{- \sin (2\lambda_0+\eta)(
\Pi(x_0)
+ \Phi(x_0) \sin \eta \sin 2\lambda_0)}{\sin(2\lambda_0-\eta)(\Pi(x_0) -
\Phi(x_0)\sin
\eta \sin 2\lambda_0)} .
\ele(zeq)
For a polynomial $f(x)$ of degree $n$ with
distinct simple roots $x_1, \dots, x_n$, one
writes
$$
f(x) = \gamma \prod_{j=1}^n ( x - x_j) = 
\gamma \prod_{j=1}^n ( \cos 2 \lambda -  \cos 2
\lambda_j) \ , \ \gamma \neq 0 .
$$
It is straight forward to see that
\begin{eqnarray}  
\eta_{q^2}f (x_k) &=& \frac{\gamma}{2^n}
\prod_{j=1}^n (qe^{2{\rm i}\lambda_k} - 
e^{2{\rm i}\lambda_j} + 
q^{-1} e^{-2{\rm i}\lambda_k}
-e^{-2{\rm i}\lambda_j }), \nonumber   \\
&=&  \frac{\gamma}{2^n}
\prod_{j=1}^n \left\{ e^{{\rm
i}(\eta+\lambda_k+\lambda_j)}  (e^{{\rm
i}(\eta+\lambda_k-\lambda_j)} \right. \nonumber \\
&{}& \qquad - \left. e^{-{\rm  i}(\eta+\lambda_k-\lambda_j)}) - 
 e^{-{\rm  i}(\eta+\lambda_k+\lambda_j)}(e^{{\rm
i}(\eta+\lambda_k-\lambda_j)}- 
e^{-{\rm i}(\eta+\lambda_k-\lambda_j)} )\right\} \nonumber \\
&=&  (-1)^n\gamma
\prod_{j=1}^n \sin (\lambda_k+\lambda_j+\eta)
\sin (\lambda_k-\lambda_j+\eta); \nonumber \\
\eta_{q^{-2}}f (x_k) &=& \frac{\gamma}{2^n}
\prod_{j=1}^n (q^{-1}e^{2{\rm i}\lambda_k} -
e^{2{\rm i}\lambda_j} +q e^{-2{\rm i}\lambda_k} 
-e^{-2{\rm i}\lambda_j}), \nonumber \\
&=& (-1)^n\gamma
\prod_{j=1}^n \sin (\lambda_k+\lambda_j-\eta)
\sin (\lambda_k-\lambda_j-\eta).\nonumber
\end{eqnarray}

Now observe that (\req(zeq)) indicates that  the roots $x_1, \ldots x_n$ of
$f(x)$ satisfy the  system of equations,
\begin{eqnarray} 
&{}& \frac{- \sin (2\lambda_k+\eta)(
\Pi(x_k)
+ \Phi(x_k) \sin \eta \sin 2\lambda_k)}{\sin(2\lambda_k-\eta)(\Pi(x_k) - \Phi(x_k)\sin
\eta \sin 2\lambda_k
)}  \nonumber \\
&{}& \qquad = \prod_{j=1}^n \frac{\sin
(\lambda_k+\lambda_j+\eta)
\sin (\lambda_k-\lambda_j+\eta)}{\sin (\lambda_k+\lambda_j-\eta)
\sin (\lambda_k-\lambda_j-\eta)} , \quad  1 \leq k
\leq n.  \nonumber
\end{eqnarray} 
In other words we arrive at the system of nonlinear equations  
\be
\frac{
\Pi(x_k)
+ \Phi(x_k)\sin \eta \sin 2\lambda_k
}{\Pi(x_k) -\Phi(x_k) \sin
\eta \sin 2\lambda_k
} = \prod_{j \neq k,  j=1}^n \frac{\sin
(\lambda_k+\lambda_j+\eta)
\sin (\lambda_k-\lambda_j+\eta)}{\sin (\lambda_k+\lambda_j-\eta)
\sin (\lambda_k-\lambda_j-\eta)} , \ \ \ 1 \leq k
\leq n ,
\ele(Bethe)
which we call the Bethe Ansatz equations
associated with $\Pi(x), \Phi(x)$. For the interest
of applications to physical problems, we shall
only consider those $q$-Sturm-Liouville problem
where the coefficients $\Pi(x), \Phi(x), r(x)$ of
(\req(SLeq)) are polynomials in $x$ with the
degrees
\be
{\rm deg}\, \Pi \ = 1+ {\rm deg}\, \Phi  \ = 2+{\rm
deg} \, r \ = N \geq 2.
\ele(deg)

\setcounter{equation}{0}

\section{Second Order Equations}

We first introduce some notations from 
\cite{An:As:Ro}, \cite{Ga:Ra}. The $q$-shifted
factorials are defined by 
\be
(a;q)_0 =1, \quad (a;q)_n = \prod_{k=1}^n(1-aq^{k-1}), 
\quad n=1, \dots, \; {\rm or}\; \infty.
\ele(shiftn)
Furthermore
\be
(a;q)_\a := \frac{(a;q)_\infty}{(aq^\a;q)_\infty}.
\ele(shiftinf)
Formula (\ref{shiftinf}) clearly holds when $\a$ is a nonnegative 
integer but is used to define the general $q$-shifted factorial 
when $\a$ is not necessarily an integer. 

Askey and Wilson \cite{As:Wi} introduced the polynomials,
\be
\phi_n(\cos \t; a) :=
(ae^{{\rm i}\t}, ae^{-{\rm i}\t};q)_n \ \equiv (ae^{{\rm i}\t};q)_n (ae^{-{\rm i}\t};q)_n \ ,
\ele(basis) 
as a basis for the space of polynomials. 
This is the most suitable basis for use here. A different basis appeared 
in the $q$-exponential 
functions in \cite{Is:Zh} where it was used to provide a $q$-analogue of 
the expansion 
of a plane wave in spherical harmonics. Clearly
\be
\phi_n(x;a) = (-2a)^n q^{n(n-1)/2}x^n + {\rm lower \; order \; terms}.
\ele(leadt)
It is straightforward to see that
$$\lim_{q \to 1}\phi_n(x;a) =  (1 - ae^{{\rm
i}\t})^n (1 -ae^{-{\rm i}\t})^n  = (1-2ax+a^2)^n.
$$ 
We may use (\req(shiftinf)) to define the more general functions 
$\{\phi_\a(x;a)\}$ by 
(\req(basis)) when $\a$ is not necessarily an integer. 
It readily follows that
\bea(ll)
\D \phi_\a(x;a) & = \frac{(1-q^\a)}{2a(q-1)}\; 
\phi_{\a-1}(x;aq^{1/2}),  
\\ 
{\cal A}_q \phi_\a(x;a) &= \phi_{\a-1}(x; aq^{1/2})
[1-aq^{-1/2}(1+q^\a)x+a^2q^{\a-1}].
\elea(adbasis)
The second formula in (\ref{adbasis}) holds when $\a=0$ provided 
that we interpret 
the product defining $\phi_\a$ in (\ref{basis}) as in (\ref{shiftinf}). 
Furthermore we have 
\be
2 {\cal A}_q \phi_\a(x;a) =  (1+q^{-\a}) \phi_{\a}(x;aq^{1/2}) + 
  (1-q^{-\a}) (1+a^2q^{2\a-1})\phi_{\a-1}(x;aq^{1/2}).
\ele(recphi) 

We shall use $\pi_j(x)$ to 
denote a generic  polynomial in $x$ of degree $j$. 
Consider a 
differential equation 
\be
\pi_2(x) y''(x) + \pi_1(x) y'(x) + \lambda y(x) =0, 
\ele(Jacobi)
where $\lambda$ is a constant. We seek a polynomial solution to (\ref{Jacobi}) 
of degree $n$. We know that one of the coefficients in $\pi_1$ or $\pi_2$ 
in not zero, hence there is no loss 
of generality in choosing it equal to 1.  Thus $\pi_1$ and $\pi_2$ contain 
four free parameters. The scaling $x \to ax+b$ of the independent variable 
absorbs two of the four parameters. The eigenvalue parameter $\l$ is 
then uniquely determined by equating coefficients of $x^n$ in (\req(Jacobi)) 
since $y$ has degree $n$.  
This reduces (\ref{Jacobi}), in general,  to a Jacobi differential equation 
whose polynomial 
solution, in general, is a Jacobi polynomial, see \cite{Sz}. Solutions also include special and limiting cases of Jacobi polynomials including the Bessel polynomials and the plynomial $x^n$.  

Next let us consider the same problem for the operator 
\be
(D_q f)(x) = \frac{f(x)-f(qx)}{x-qx}.
\ele(qlin)
Consider the operator equation 
\be
\pi_2(x) D_q^2y(x) + \pi_1(x) D_q y(x) + \lambda y(x) =0.
\ele(q-Jacobi)
Here under the same assumptions on $\pi_1$ and $\pi_2$  
one easily finds out that the only scaling allowed on $x$ is  $x \to ax$, 
hence one of the coefficients in $\pi_1$ and $\pi_2$ is chosen as 1 and the 
remaining coefficients of  
$\pi_1$ and $\pi_2$ constitute  three free parameters. 
 The general polynomial solution to 
 (\ref{q-Jacobi}) is the big $q$-Jacobi polynomial which contains 
three free parameters, \cite{Ko:Sw}, \cite{Ga:Ra}, \cite{An:As:Ro}. 

We next consider the\aw case
\be
\pi_2(x) \D^2 y(x) + \pi_1(x) {\cal A}_q \D y(x) 
+ \lambda y(x) =0.
\ele(awe) 
In this case one  can not perform any scaling on
$x$, so apart from assuming that  one of the
coefficients in $\pi_1$ and $\pi_2$ is unity, we have four
free parameter,  namely the remaining
coefficients in $\pi_2$ and the coefficients in  
$\pi_1$. This is the case of the\awp where 
\bea(lll)
\pi_2(x) &= & -q^{-1/2}\left[
2(1+\sigma_4)x^2-(\sigma_1+\sigma_3)x 
-1+\sigma_2-\sigma_4\right], \\
\pi_1(x) &= & 2\left[ 2(\sigma_4-1) x +\sigma_1- \sigma_3\right]/(1-q),
\elea(q-Jacobi2)
where $\sigma_j$ is the $j$th elementary symmetric function of parameters 
$a_1, a_2, a_3, a_4$. In order to solve (\req(awe)) for $y$ we 
expand  $y$ in the\aw basis $\{\phi_n(x;a_1)\}$ and find 
\be
\l = -4q(1-q^{-n})(1-\sigma_4q^{n-1})(1-q)^{-2}.
\ele(AWl) 
With the above choice for $\lambda$ the polynomial 
solution to (3.3) is unique and
is given by an Askey-Wilson polynomial of 
degree $n$, see for example \cite{Br:Ev:Is}.

At this stage one wonders whether replacing $\pi_2, \pi_1$ and $\l$ 
by $\pi_{k+1}, \pi_k$, and $\pi_{k-1}$ in 
equations (\req(Jacobi)), (\req(q-Jacobi))  and (\req(awe)) 
lead to more general orthogonal polynomials. Grunbaum and Haine
\cite{Gr:Ha}  proved that the only orthogonal
polynomial solutions to (\req(awe)) after the replacements 
$(\pi_2, \pi_1,\l) \to (\pi_{k+1}, \pi_k, \pi_{k-1})$ are  the\awp
or special and limiting cases of them.  Ismail \cite{Is03} showed that the same conclusion holds without assuming orthogonality. This generalizes earlier work of 
Hahn, and 
Bochner who proved that the $q$-Jacobi polynomials, and the Jacobi polynomials 
are the only polynomial solutions to (\req(q-Jacobi)) and 
(\req(Jacobi)), respectively with the above replacements.

\setcounter{equation}{0}

\section{Multiparameter Cases and Bethe Ansatz
Equations for the XXZ Model} For $2N$ complex numbers $a_j, 1
\leq j
\leq 2N$, we denote 
$$
\vec{a} = (a_1, \ldots, a_{2N}), 
$$
and $\sigma_j$ the $j$-th elementary symmetric
function of $a_i$'s for $0 \leq j \leq 2N$,  with 
$\sigma_0 := 1$. We define the weight
function $w(x, \vec{a})$,
\be
w(x, \vec{a} ) : = \frac{(e^{{\rm i}N \theta},
e^{-{\rm i}N \theta}; q^{\frac{N}{2}})_\infty
}{\sin \frac{N\theta}{2}\prod_{j=1}^{2N}(
a_j e^{{\rm i}\theta}, a_j
e^{-{\rm i}\theta}; q)_\infty},
\ele(qwt)
which can also be written in the following form,
\be
w(x, \vec{a})= \frac{2{\rm i} e^{\frac{-{\rm i}N
\theta}{2} }  (e^{{\rm i}N
\theta}, q^{\frac{N}{2}} e^{-{\rm i}N \theta};
q^{\frac{N}{2}})_\infty }{\prod_{j=1}^{2N}(
a_j e^{{\rm i}\theta}, a_j e^{-{\rm i}\theta};
q)_\infty} 
= \frac{-2{\rm i} e^{\frac{{\rm i}N \theta}{2}
}  (q^{\frac{N}{2}} e^{{\rm i}N
\theta},  e^{-{\rm i}N \theta};
q^{\frac{N}{2}})_\infty }{\prod_{j=1}^{2N}(
a_j e^{{\rm i}\theta}, a_j e^{-{\rm i}\theta};
q)_\infty}.  
\ele(wN)
With $w(x) = w(x, \vec{a}), p(x)= w(x, q^{1/2}\vec{a})$ in (\req(SL)), 
we shall consider the following equations,
\be
\frac{1}{w(x, \vec{a})} {\cal D}_q ( ({w(x,
q^{1/2}\vec{a})} {\cal D}_q ) y ) (x) = r(x)
y(x), \quad r(x) \in \CZ[x] \ , \ {\rm deg} \ 
r = N-2.
\ele(SLe)
With the notation 
\be 
\Pi(x; \vec{a}) = \frac{1}{w(x, \vec{a})}
{\cal A}_q w(x,
q^{1/2}\vec{a}) \ , \ \ 
\Phi(x; \vec{a}) =  \frac{1}{w(x, \vec{a})}
{\cal D}_q w(x,
q^{1/2}\vec{a}),
\ele(wPiPhi)
the equation in (\req(SLe)) becomes 
\be
\Pi(z; \vec{a}) \D^2 y + \Phi(z; \vec{a}) {\cal A}_q \D y = r(x) y. 
\ele(SLe2)
Note that for $N=2$, $r(x)= r \in \RZ$ and $a_j ,
q
\in
\RZ$ with
$|a_j| <1, 0 < q<1$, the weight function
$w_{\vec{a}}(x)$ is a positive function on $(-1,
1)$, and the solutions of the equation (\req(SLe))
are the Askey-Wilson polynomials which are
orthogonal in $L^2((-1,
1); (1-x^2)^{-1/2})$, see e.g.
\cite{Br:Ev:Is}. 

\begin{thm} \label{thm:PiPhi} The functions 
$\Pi(x; \vec{a}), \Phi(x;\vec{a})$ are 
polynomials of $x$ of degree $N, N-1$
respectively, and have 
the following explicit forms,
\begin{eqnarray}
\label{Pia}
\Pi(x; \vec{a}) &=& -q^{-N/4} \{ (-1)^N
\sigma_N +
\sum_{l=0}^{N-1}(-1)^l ( \sigma_l + \sigma_{2N-l})
T_{N-l}(x) \},  \\      \label{Phia}
\Phi(x; \vec{a}) &=&
\frac{2q^{-N/4}}
{q^{1/2}-q^{-1/2}}
\sum_{l=0}^{N-1}(-1)^l ( \sigma_l - \sigma_{2N-l})
U_{N-l-1}(x). 
\end{eqnarray}
Conversely , for given polynomials $\Pi$ and $\Phi$ of degrees $N$ 
and $N-1$ respectively, there is a unique $2N$-element set 
$\{a_j: 1 \le j \le 2N\}$ such that $\Pi(x) = \Pi(x, \vec{a})$ and 
$\Phi(x) =  \Phi(x; \vec{a})$ and (\req(wPiPhi)) holds.  
\end{thm}
{\it Proof:} By the definition of $w(x,
\vec{a})$, also the forms in (\req(wN)) when
applying $\eta_q, \eta_{q^{-1}}$, $\Pi(x; \vec{a})$ 
has the following expression, 
$$ 
\frac{{\rm i} q^{\frac{-N}{4}} \sin
\frac{N\theta}{2}\prod_{j=1}^{2N}( a_j e^{{\rm
i}\theta}, a_j e^{-{\rm i}\theta};
q)_\infty}{(e^{{\rm i}N
\theta}, e^{-{\rm i}N \theta};
q^{\frac{N}{2}})_\infty }
\left[ \frac{  e^{\frac{-{\rm i}N
\theta}{2} }  (q^{\frac{N}{2}}e^{{\rm i}N
\theta},  e^{-{\rm i}N \theta};
q^{\frac{N}{2}})_\infty }{\prod_{j=1}^{2N}(
a_j q e^{{\rm i}\theta}, a_j e^{-{\rm
i}\theta}; q)_\infty} 
- \frac{ e^{\frac{{\rm i}N \theta}{2} }  ( e^{{\rm i}N
\theta},  q^{\frac{N}{2}}e^{-{\rm i}N \theta};
q^{\frac{N}{2}})_\infty }
{\prod_{j=1}^{2N}( a_j e^{{\rm i}\theta}, a_j q
e^{-{\rm i}\theta}; q)_\infty}\right],
$$
which implies
\begin{eqnarray} 
\Pi(x; \vec{a}) 
&=& {\rm i} q^{\frac{-N}{4}} \sin
\frac{N\theta}{2} \left[ \frac{e^{\frac{-{\rm i}N
\theta}{2} }}{1-e^{{\rm i}N\theta}}
\prod_{j=1}^{2N} (1-a_je^{{\rm i}\theta})- 
\frac{e^{\frac{{\rm i}N
\theta}{2} }}{1-e^{-{\rm i}N\theta}}
\prod_{j=1}^{2N} (1-a_je^{-{\rm i}\theta}) \right] \nonumber   \\
& =& \frac{-q^{-N/4}}{2}  \left[e^{-{\rm i}N \theta}
\prod_{j=1}^{2N} (1-a_je^{{\rm i}\theta})+
e^{{\rm i}N \theta }
\prod_{j=1}^{2N} (1-a_je^{-{\rm i}\theta})\right] \nonumber   \\     
&=&  \frac{-q^{\frac{-N}{4}}}{2}  
\sum_{l=0}^{2N}(-1)^l \sigma_l (e^{{\rm
i}(l-N)\theta}+ e^{{\rm i}(N-l)\theta}) = -
q^{\frac{-N}{4}} 
\sum_{l=0}^{2N}(-1)^l \sigma_l \cos (N-l)\theta \nonumber  \\
& =& - q^{\frac{-N}{4}} \{ (-1)^N \sigma_N +
\sum_{l=0}^{N-1}  (-1)^l (\sigma_l + \sigma_{2N-l})
T_{N-l}(x) \}. \nonumber
\end{eqnarray}
By the same method, we have 
\begin{eqnarray}  
\Phi(x; \vec{a}) &=& \frac{2  q^{\frac{-N}{4}} \sin
\frac{N\theta}{2} }{
(q^{1/2}- q^{-1/2})
\sin \theta  }
\left[ \frac{e^{\frac{-{\rm i}N
\theta}{2} }}{1-e^{{\rm i}N\theta}}
\prod_{j=1}^{2N} (1-a_je^{{\rm i}\theta})+ 
\frac{e^{\frac{{\rm i}N
\theta}{2} }}{1-e^{-{\rm i}N\theta}}
\prod_{j=1}^{2N} (1-a_je^{-{\rm i}\theta}) \right] \nonumber \\
\label{Phiexp} &=& \frac{{\rm i}  q^{\frac{-N}{4}} }{
(q^{1/2}- q^{-1/2})
\sin \theta  }
\left[e^{-{\rm i}N
\theta} 
\prod_{j=1}^{2N} (1-a_je^{{\rm
i}\theta})- e^{{\rm i}N \theta}
\prod_{j=1}^{2N} (1-a_je^{-{\rm i}\theta}) \right]  
\\
&=& \frac{{\rm i}  q^{\frac{-N}{4}} }{
(q^{1/2}- q^{-1/2})
\sin \theta  } 
\sum_{l=0}^{2N}(-1)^l \sigma_l(e^{{\rm
i}(l-N)\theta}- e^{{\rm i}(N-l)\theta}) \nonumber \\
&=&  \frac{2 
q^{\frac{-N}{4}} }{ (q^{1/2}-
q^{-1/2})  } 
\sum_{l=0}^{2N}(-1)^l \sigma_l \frac{\sin
(N-l)\theta}{\sin \theta}\nonumber\\
& =& \frac{2 
q^{\frac{-N}{4}} }{ (q^{1/2}-
q^{-1/2})  } 
\sum_{l=0}^{N-1}(-1)^l (\sigma_l-
\sigma_{2N-l})U_{N-l-1}(x).  \nonumber 
\end{eqnarray}
To see the converse statement, given $\Pi$ and $\Phi$ we 
expand them in Chebyshev polynomials of the first and second kinds 
respectively, then define $\sigma_0$ by $\sigma_0 =1$ and $\sigma_N$ 
by $(-1)^{N+1} q^{N/4}$ times the constant  term in the expansion (\req(Pia)) in terms of 
Chebyshev polynomials. Then define 
the remaining $\sigma$'s through finding $\sigma_l \pm \sigma_{2N-l}$ from  
the coefficients in $\Pi$ and $\Phi$ in (\req(Pia)) and (\req(Phia)).  
$\Box$ 

It is important to note that Theorem \ref{thm:PiPhi} gives a constructive way of 
identifying the parameters $a_1, \dots a_{2N}$ and $w(x; \vec{a})$ 
from the functional equation (\req(SLe2)).

\begin{thm} \label{thm:Betheg} 
Let $a_l =q^{-s_l} = e^{-2 i \eta s_l}$ for $1 \leq l \leq 2N$.
The Bethe Ansatz equations {\rm(\req(Bethe))} associated 
with the polynomials 
$\Pi(x; \vec{a}), \Phi(x; \vec{a})$ have the
form $(\req(gBA))$.
\end{thm}
{\it Proof:} By Theorem \ref{thm:PiPhi}, (indeed
in its proof), we have
\begin{eqnarray} 
\Pi(x; \vec{a}) 
&=& \frac{-q^{\frac{-N}{4}} }{2}  \left[
e^{-2{\rm i}N
\lambda_k}
\prod_{j=1}^{2N} (1-a_je^{2{\rm i}\lambda_k})+
e^{2{\rm i}N
\lambda_k }
\prod_{j=1}^{2N} (1-a_je^{-2{\rm i}\lambda_k}) \right]  \nonumber \\
\Phi(x_k; \vec{a}) \sin \eta \sin 2\lambda_k
& =&
 \frac{
q^{\frac{-N}{4}}  }{ 2 } \left[e^{-2{\rm i}N
\lambda_k} 
\prod_{j=1}^{2N} (1-a_je^{2{\rm
i}\lambda_k})- e^{2{\rm i}N \lambda_k}
\prod_{j=1}^{2N} (1-a_je^{-2{\rm i}\lambda_k})\right], \nonumber 
\end{eqnarray}
hence 
\begin{eqnarray}
\Pi(x_k; \vec{a})  + \Phi(x_k; \vec{a}) \sin
\eta \sin 2\lambda_k
 &=& -q^{\frac{-N}{4}}
e^{2{\rm i}N
\lambda_k }
\prod_{j=1}^{2N} (1-a_je^{-2{\rm i}\lambda_k})\nonumber \\
 &=&  -q^{\frac{-N}{4}}
e^{-2{\rm i}N
\lambda_k }
\prod_{j=1}^{2N} (e^{2{\rm i}\lambda_k}-a_j), \nonumber \\
\Pi(x_k; \vec{a}) - \Phi(x_k; \vec{a}) \sin
\eta \sin 2\lambda_k
  & =& -
q^{\frac{-N}{4}}  
e^{-2{\rm i}N
\lambda_k} 
\prod_{j=1}^{2N} (1-a_je^{2{\rm
i}\lambda_k}). \nonumber 
\end{eqnarray}
By substituting $a_j=q^{-s_j}$ in the above
formula, the result of this theorem follows from 
(\req(Bethe)).
$\Box$ \par \vspace{.2in} 
If $|q| > 1$ replace $q$ by $1/p$, use the invariance of 
${\cal D}_q$ and ${\cal A}_q$ under $q \to q^{-1}$ to rederive
Theorem 4.2 with $q$ replaced by $1/p$. This covers the case $|q| >
1$. The cases $|q| =1$ and in particular the cases when $q$ is a
root of unity do not seem to be amenable to the techniques
developed here.  

The Bethe Ansatz equations (\req(Bethe))
describe the relations of roots $x_j, 1 \leq j
\leq n$, of a polynomial $f(x)$ of degree $n$ in
the $q$-Sturm-Liouville problem (\req(SLe)). It 
is important to note that for given $\Pi$ and
$\Phi$ in  (\req(SLeq)) with  
deg$\Pi = N$ and deg$\Phi = N-1$, there are $2N$ complex numbers, $a_1, a_2, 
\dots, a_{2N}$, such that $\Pi(x) = \Pi(x; \vec{a})$ and  
$\Phi(x) = \Phi(x; \vec{a})$.
For a positive integer $N$ and all
$a_j= q^{-s}$ with $s \in
\frac{1}{2}\ZZ_{\geq 0}$, the equation
(\req(Bethe)), or  (\req(gBA)) become  (\req(BetheXXZ))
which is the Bethe Ansatz equations for the  spin $s$
XXZ model  of a size
$2N$ with the open (Dirichlet) boundary
condition \cite{S}.  This correspondence shows that identifying the spectrum of 
XXZ spin chain is related to spectral problems of a $q$-Sturm-Liouville equation.

For $N=2$ , we have
$\Pi(x; \vec{a}) = \pi_2(x)$ and $
\Phi(x; \vec{a}) = \pi_1(x) $ in
(\req(q-Jacobi2)). 
In the special case
$$
a_1=-a_2 = e^{-{\rm i}s}, \quad 
a_3=-a_4 = e^{-{\rm i}(s+\eta)}, 
$$
the polynomial solution to (\req(SLe)) is a $q$-ultraspherical 
polynomial of degree $n$. In this case
(\req(gBA)) becomes
$$
\frac{\sin (2\lambda_k +s) \sin (2\lambda_k
+s+\eta)     }{\sin (2\lambda_k -s)
\sin (2\lambda_k -s-\eta)}  =
\prod_{j
\neq k,  j=1}^n \frac{\sin
(\lambda_k+\lambda_j-\eta)
\sin (\lambda_k-\lambda_j-\eta)}{\sin
(\lambda_k+\lambda_j+\eta)
\sin (\lambda_k-\lambda_j+\eta)} , \ \ \ 1 \leq k
\leq n.
$$
For $N=2$ with $s_j <0$ for all $j$,  $y(x)$ is an
Askey-Wilson polynomial of degree $n$. In fact,
one has the following result.
\begin{thm} \label{AWpoly}
Let $N=2$ and $\eta = i\zeta$, $\zeta >0$. Then for all $n$ the system 
(\req(gBA)) has 
a unique solution provided 
that $s_j < 0$, $1 \le j \le 4$. Furthermore 
all the $\l$'s are in $(0, \pi/2)$. 
\end{thm}
\noindent{\it Proof}.  Let $y$ be a polynomial
of degree $n$ with  zeros $\cos (2\l_j)$,  $1\le
j \le n$. We know that (\req(gBA)) implies the  
validity of (\req(awe)) for $x = \cos
(2\l_j)$.   Here
$\pi_1$ and $\pi_2$ are as in (\req(awe)).
With
$\l$ in (\req(q-Jacobi)) chosen   as in
(\req(AWl)) the left-hand side of
(\req(awe)) is a polynomial in $x$ of  
degree $n-1$ and vanishes at $n$ points. Hence
(\req(awe)) must hold for   all
$x$, and
$y$ must be an
\aw polynomial of   degree $n$.  Since the\awp are
orthogonal on  $[-1, 1]$,  all their zeros are
in $(-1, 1)$. 
$\Box$ \par \vspace{.2in} 

For large $n$ the distribution of the $\cos (2\l_j)$'s follows is an arcsine 
distribution. This follows from general theory of orthogonal polynomials 
since in this case $\int_{-1}^1 \ln w(x) (1-x^2)^{-1/2}dx$ is finite \cite{Ne1}, 
\cite{Ne2}.  
Note that the proof of  Theorem \ref{AWpoly} used the
orthogonality of the\awp in an essential  way.

Using arguments that go back to Heine,  one can proved that for given
polynomials $\Pi$ and $\Phi$ of degree 
$N$ and $N-1$ 
respectively, there exist at most ${N+n-2 \choose N-2}$ 
choices of $r(x)$ such that 
(\req(SLeq)) has a polynomial solution of degree $n$. This shows that for $N > 2$  the solution to
(\req(gBA)), without additional assumptions, may
not be unique. 

For $a_j = q^{-s_j}$ the monic\awp satisfy the 
recurrence relation 
\begin{eqnarray}
xp_n(x) = p_{n+1}(x) + (\cos 2 \eta + 2 A_n + 2 C_n) p_n(x) + 4 A_{n-1}C_np_{n-1}(x),
\end{eqnarray}   
where
\bea(ll)
A_n &=\frac{\sin ((n-1 - \sigma_1^{\prime})\eta)
 \prod_{j=2}^4 \sin ((n -s_1 -  s_j) \eta)}{\sin((2n-1- \sigma_1^{\prime})\eta)
 \sin((2n- \sigma_1^{\prime})\eta)}, \\
 &{} \\
 C_n &=  \frac{\sin (n\eta)\prod_{2\le j <k\le 4} \sin((n-1-s_j-s_k)\eta)}  
 {\sin((2n-1- \sigma_1^{\prime})\eta)\sin((2n-2- \sigma_1^{\prime})\eta)},
\elea(ACn)
and $\sigma_j^{\prime}$ is 
the $j$th elementary symmetric function of the $s_k$'s, (see (3.1.4)-(3.1.5) in \cite{Ko:Sw}). From the form of $A_n$ 
and $C_n$ in (\req(ACn)) it is clear that the
positivity condition 
$A_{n-1}C_n > 0$ may not hold for all $n$. 
\begin{thm}
Assume that $\eta$ and all the $s_j$'s are real 
and that $A_0C_1 > 0$. Let 
\begin{eqnarray*}
M = {\rm Max}\{n: A_{k-1}C_k > 0,\;  {\rm for}\; 
1\le k \le n\}.
\end{eqnarray*}
Then $(\req(gBA))$ has a unique solution for 
$0 \le n \le M$ and the $\l$'s are 
all real and simple. 
\end{thm}
{\it Proof}. We have a family of $M+1$ (which may be $=\infty$) orthogonal
polynomials and their zeros are all real and simple. It is known that in this
case (4.5) has only one polynomial solution \cite{At:Su}.
$\Box$ \par \vspace{.2in} 

\noindent
{\bf Remark}: It is important to emphasize that when the orthogonality condition
$A_{n-1}C_n > 0$ is violated, then the Askey-Wilson polynomials continue to
satisfy (4.5), hence their zeros solve the Bethe Ansatz equations but we can no
longer guarantee the reality or the simplicity of the zeros of the Askey-Wislon
polynomials.
$\Box$
\setcounter{equation}{0}

\section{Singularities and Expansions}  
In this section we consider different regimes. We assume $0 < q < 1$ so ${\rm i} \eta < 0$ only in (\ref{InPr})-(\ref{SLevp}) below.
Recall the inner product  
associated with the Chebyshev weight  $(1-x^2)^{-1/2}$ on 
 $ (-1, 1)$, namely
\begin{eqnarray}
\label{InPr}
<f, g> := \int_{-1}^1 f(x)\; \overline{g(x)}\; \frac{dx}{\sqrt{1 - x^2}}.
\end{eqnarray}
For $0 < q < 1$, one observes that the definition (\req(Ddef))  requires $\breve{f}(z)$ to 
be defined for $|q^{\pm1/2}z|=1$ as well as for $|z|=1$. In particular 
${\cal D}_q$
is well-defined on $H_{1/2}$, where
\begin{eqnarray}
\label{Hp}
H_{\nu} := \{f: f((z+1/z)/2) \mbox{ is analytic for } q^{\nu} \le |z| \le
q^{-\nu} \}.
\end{eqnarray} 
Brown, Evans, and Ismail \cite{Br:Ev:Is} proved the  integration by parts 
formula
\begin{eqnarray}
\label{IBP}
<{\cal D}_q\, f, g> & =& \frac{\pi\,\sqrt{q}}{1-q} 
\left[ f(\frac{1}{2}(q^{1/2}+q^{-1/2}))
 \overline{g(1)} - f(-\frac{1}{2}(q^{1/2}+q^{-1/2})) 
 \overline{g(-1)}\right]\\
&\mbox{}& \quad
- <f, \sqrt{1-x^2}
\,{\cal D}_q(g(x)\, (1-x^2)^{-1/2})>, 
\nonumber
\end{eqnarray}
for $f$, $g \in H_{1/2}$. They also proved that if $w(x) > 0$ on $(-1,1)$, 
and $y \in H_1$, $p  \in H_1$ then the eigenvalues of the 
$q$-Sturm-Liouville problem, 
\begin{eqnarray}
\label{SLevp}
\frac{1}{w(x)} {\cal D}_q \left( p(x) {\cal D}_q  y(x)\right)
= \l y(x),
\end{eqnarray}
are real and the eigenfunctions corresponding to distinct eigenvalues are 
orthogonal  with respect to $w$. This implies the following theorem.
\begin{thm}
Assume that $0 < q < 1$, $|a_j| < q^{1/2}, \ 1 \leq j \leq 4,$ and $w(x; \vec{a}) > 0$ on $(-1,1)$ where $\vec{a} = (a_1, a_2, a_2, a_4)$. Then the eigenvalues of 
$$
\frac{1}{w(x, \vec{a})} {\cal D}_q \left( w(x,q^{1/2}\vec{a} ) {\cal D}_q  y(x)\right)
= \l y(x),
$$
are real and simple, and the eigenfunctions are mutually orthogonal on $(-1, 1)$ with respect to the weight function $w(x;\vec{a})$. 
\end{thm}
{\it Proof}. The condition $|a_j| < q^{1/2}$, $1 \le j \le 4$ ensures that 
$p(x) = w(x; q^{1/2}\vec{a})$ is in $H_1$ and the theorem follows from 
\cite{Br:Ev:Is}. $\Box$

We now discuss the concept of a regular singular point of the equation  
(\req(SLe)), which is the general form of a second order equation in the\aw 
operator with polynomial coefficients as we say in Theorem \ref{thm:PiPhi} for a general $q$ and $N$.

First recall that the concept of singularities of differential equations is 
related to the analytic properties of the solutions in  a neighborhood of the 
singularities. We have no geometric way to describe the corresponding 
situation for equations like (\req(SLe)). In the 
present set up the analogue of a function analytic in a neighborhood of 
a point $(a+a^{-1})/2$ is a function which has a convergent series 
expansion of the form $\sum_{n=0}^\infty c_n \phi_n(x;a)$. We have no other 
characterization of these $q$-analytic functions. 

In the case of second order differential equations the singularities 
are the zeros of the coefficient of $y''$ and the regular singular points 
$\zeta_j$'s are precisely those points 
where one can construct a 
series solution to the differential equation of the form 
$\sum_{n=0}^\infty c_n (z-\zeta_j)^{\a_j+n}$ in a neighborhood of $z = \zeta_j$. 
A closer examination of the equation (\req(SLe)) in the form 
(\req(SLe2)) 
reveals that one can formally expand a solution $y$ as $\sum_{n=0}^\infty 
y_n \phi_n(x;a)$, substitute the series expansion in (\req(SLe2)) and 
recursively compute 
the coefficients $y_n$ provided that $a$ is not among the $2N$ parameters 
$\{\zeta_1, \dots, \zeta_{2N}\}$, where $\zeta_j = (a_j+ a_j^{-1})/2$, 
and possibly the points $0$ and $\infty$, which we 
do not know how to handle. This indicates that what plays the role
of singular points  of (\req(SLe2)) are $\zeta_1, \dots \zeta_{2N}$
in addition to
$x =0, x= 
\infty$.  
Around the singular point $x = \zeta_j$, we  shall expand functions in the 
set $\{\phi_{n+\a}(x; a_j)\}$, where $\phi_\a(x; a)$ is as in (\req(basis)).
We let 
\be
y(x) = \sum_{n=0}^\infty y_n \; \phi_{n+\a}(x; a_1),
\ele(expansion)
and observe that $r(x)  \phi_{n+\a}(x; a_1)$ is a linear 
combination of  $\{\phi_{m+\a}(x; a_1): n \le m \le n+ N-2\}$. Furthermore 
we note that 
(\req(shiftinf)), (\req(basis)), and (\req(wPiPhi)) imply 
\begin{eqnarray}
&{}& 
\frac{1}{w(x; \vec{a})}\D \left(w(x; q^{1/2}\vec{a}) \D 
\phi_{n+\a}(x; a_1)\right) \nonumber \\
&{}& \quad =
\frac{1-q^{n+\a}}{2a_1(q-1)w(x; \vec{a})} 
\D \left(w(x; q^{1/2}\vec{ a}) \phi_{n+\a-1}(x; q^{1/2} a_1)\right)\nonumber \\
&{}& \quad = \frac{1-q^{n+\a}}{2a_1(q-1)w(x; \vec{a})} 
\D \left(w(x; a_1q^{n+\a-1/2}, a_2, \dots, a_{2N})\right)  \nonumber \\
&{}& \quad = \frac{(1-q^{n+\a})w(x; a_1q^{n+\a-1}, a_2, \dots, a_{2N})}{2a_1(q-1)w(x; \vec{a})}\, 
\Phi(x; a_1q^{n+\a-1}, a_2, \dots, a_{2N}) \nonumber \\
&{}& \quad = \frac{1-q^{n+\a}}{2a_1(q-1)} \, \Phi(x; a_1q^{n+\a-1}, a_2, \dots, a_{2N})\, 
 \phi_{n+\a-1}(x; a_1). 
\nonumber 
\end{eqnarray}
The substitution of  the expansion (\req(expansion)) for $y$ in  
(\req(SLe)), reduces the left-hand side of (\req(SLe)) to 
\begin{eqnarray}
\sum_{n=0}^\infty \frac{1-q^{n+\a}}{2a_1(q-1)} \,\Phi(x; a_1q^{n+\a-1}, a_2, \dots, a_{2N})\, 
 \phi_{n+\a-1}(x; a_1) y_n. \label{regex}
\end{eqnarray}
Note that the smallest subscript of a $\phi$ in $r(x) y(x)$ in 
(\req(SLe2)) is $\a$. On the other hand  (\ref{regex})
implies that $\phi_{\a-1}$ appears on the left-hand 
side of (\req(SLe)). 
Thus the coefficient of $\phi_{\a-1}(x; a_1)$ must be zero. To determine this 
coefficient we set  
\begin{eqnarray}
\Phi(x; q^{\a-1}a_1, a_2, \dots, a_{2N}) = \sum_{j=0}^{N-1} 
d_j(q^\a)\, \phi_j(x; a_1q^{\a-1}),
\end{eqnarray}
and after making use of $\phi_n((a+a^{-1})/2; a) = \delta_{n,0}$ we find 
\begin{eqnarray}
d_0(q^\a) = \Phi((a_1q^{\a-1}+ a_1^{-1}q^{1-\a})/2; q^{\a-1}a_1, a_2, 
\dots, a_{2N}).   \nonumber
\end{eqnarray}
Thus the vanishing of the coefficient of $\phi_{\a-1}(x; a_1)$ on the 
left-hand side of (\req(SLe)) implies the vanishing of $(1-q^\a) 
d_0(q^\a)$, that is 
\begin{eqnarray}
\label{indicial}
(1-q^\a) \Phi((a_1q^{\a-1}+ a_1^{-1}q^{1-\a})/2; q^{\a-1}a_1, a_2, \dots, a_{2N})
= 0. 
\end{eqnarray}
\begin{thm}
Assume $|a_j| \le 1$, for all $j$.  Then  the only solution(s) of (\ref{indicial}) 
are given by $q^\a = 1$, or $q^\a = q/(a_1a_j)$, $ j =2, \dots, 2N$. 
\end{thm}


\noindent{\it Proof}. From (\req(indicial)) it is 
clear that $q^\a =1$ is a solution.  
With  $x = (a_1q^{\a-1}+ a_1^{-1}q^{1-\a})/2$ as in 
(\req(indicial)) we find 
$e^{i\t} = a_1q^{\a-1},$ or $a_1^{-1} q^{1-\a}$. In the former case,  
$2 {\rm i}\sin \t = a_1q^{\a-1}- a_1^{-1}q^{1-\a}$, hence (\req(indicial)) 
and (\req(Phiexp))  imply
\begin{eqnarray}
\frac{2 {\rm i}(1-a_1^2q^{2\a-2})}{a_1q^{\a-1}- q^{1-\a}/a_1} 
\;  \prod_{j=2}^{2N}(1- a_1a_jq^{\a-1}) = 0,   \nonumber 
\end{eqnarray}
which gives the result. On the other hand if $e^{i\t} = 
q^{1-\a}/a_1$, then we reach the same solutions via (\req(Phiexp)). 
$\Box$ \par \vspace{.2in}

Ismail and Stanton  \cite{Ism:Sta} used two  bases in addition to
$\{\phi_n(x;a)\}$ for polynomial expansions. Their bses are 
\begin{eqnarray}
\r_n(\cos \t) &:=& (1+e^{2i\t})(-q^{2-n}e^{2i\t};q^2)_{n-1}
e^{-in\t},
\\
\phi_n(\cos \theta) &: =& (q^{1/4}e^{i\theta},
q^{1/4}e^{-i\theta};q^{1/2})_n.
\end{eqnarray}
They satisfy 
\begin{eqnarray}
{\cal D}_q \r_n(x) &=& 2q^{(1-n)/2}\frac{1-q^n}{1-q}\,
\r_{n-1}(x),\\
{\cal D}_q \phi_n(x) &=& -2q^{1/4}\,\frac{1-q^n}{1-q}
\, \phi_{n-1}(x)
\end{eqnarray}
One can also seek solutions of second order operator equations by 
expanding solutions in the above polynomial basis. This will be the
subject of future work.

Atakishiyev and Suslov \cite{At:Su} studied certain expansions of 
solutions of a very special nonhomogenous equation 
corresponding to (\req(awe)), that is $N=2$, with special value of $\l$. 
They did not however investigate any concept of singularities, nor they have  
observed the general structure of expanding the polynomial coefficients in 
Chebyshev polynomials. Their Wronskian is inadequate because
according to their definition, the Wronskian of two polynomials
is not a polynomial.  Askey and Wilson  wrote down the functional
equation (\req(awe)) and identified it  as the equation satisfied by
the\aw polynomials.  Of course when one works only with the case $N
=2$, as in \cite{At:Su}, one  is bound to miss the complications and the
elegance  of the case of  general  $N$.

\setcounter{equation}{0}

\section{Bethe Ansatz Equations for the XXX Model and Wilson
Operators}
In this section we consider the
problems arisen from the discussion in the
previous sections when $q$ tends to
one, the corresponding
Sturm-Liouville problem connecting with the Bethe
Ansatz equations of the Heisenberg XXX spin chain. We write
the variable
$z$ in previous sections in the form,
$$
z = e^{{\rm i}\theta} = q^{-{\rm i}y}, 
$$
and again the parameters, $
a_j = q^{-s_j}$. It is known that as $q$ tends
to 1, the transformation $\breve{f}(z)$ and 
the operators 
$\eta_{q^{\pm 1}}, {\cal D}_q, {\cal A}_q$ become
$\check{f}(y), \eta_\pm, W, A$, respectively, where
$$
\begin{array}{rl}
\check{f}(y) &: = f(x)  \ \ \ {\rm with } \ x= y^2 \ \ ;
\\ (\eta_\pm f) (x) &: = \check{f} (  y \pm
\frac{\rm i}{2} )
\ , \\ 
(W f) (x) &: = \frac{1}{2y{\rm i}}
(\eta_+ f -\eta_-f)(x) 
\ ,
\\
(A f) (x) &: = \frac{1}{2}(\eta_+ f +
\eta_- f)(x).
\end{array}
$$ 
The above divided difference operator $W$ is called the
Wilson operator \cite{W}. We have the relation
$$
W(fg) = (Wf) (Ag) +( Af)(Wg).
$$
Analogous to the $q$-Sturm-Liouville problem
(\req(SL)), we consider the following difference
equation
$$
\frac{1}{w(x)} W \left( p(x) W  f\right) (x)
= r(x) f(x) \ ,
$$
which is equivalent
to the Sturm-Liouville problem in the form, 
\be   
\Pi(x) W^2  f
(x) +\Phi(x)(A W f)(x) = r(x) f(x), 
\ele(SLeqW)
where  $\Pi, \Phi$ are the functions defined by 
\be
\Pi(x)= \frac{1}{w(x)} A p (x), \quad \Phi(x)= \frac{1}{w(x)} W p(x) .
\ele(wPi)
For our purpose with reason which will be
clearer later on, we seek polynomial solutions $f(x)$ to (\req(SLeqW)), but
$\Pi (x),
\Phi (x), r(x)$ are rational functions with the same degree constraints as in (\req(deg)).
From the relations
$$
\begin{array}{ll}
AWf (x) & = \frac{1}{{\rm i} (4 y^2+1)} \{ (y-
\frac{\rm i}{2}) \eta_+^2 - (y+ \frac{\rm i}{2})
\eta_-^2 + {\rm i} 
\} f (x) ; \\ 
W^2 f (x) & = \frac{-1}{y (4 y^2+1)}
\{
( y -
\frac{\rm i}{2})
\eta_+^2 +(  y + \frac{\rm
i}{2})
\eta_-^2 -2 y \}f 
(x),
\end{array}
$$
it follows that if $f(x_0)=0$ for $x_0=y_0^2$, then 
\be  
\left( \frac{\eta_+^2 f}{\eta_-^2f} \right) (x_0) = \frac{- ( 
y_0 +
\frac{\rm i}{2})(\Pi(x_0)- \Phi(x_0)  y_0{\rm i}
 ) }{( y_0 -
\frac{\rm i}{2})( \Pi(x_0) + \Phi(x_0)  y_0{\rm i}  
)  } . 
\ele(etar)
For a degree $n$ polynomial $f(x)$  with roots
$x_j$ for $1 \leq j \leq n$, and let $x=y^2$, $x_j = y_j^2$, that is  
$$
f(x) = \gamma \prod_{j=1}^n ( x - x_j) = 
\gamma \prod_{j=1}^n (y^2 -  y_j^2), \ \gamma
\neq 0.
$$
Then
$$
\eta_\pm ^2 f (x_k) = \gamma
\prod_{j=1}^n (y_k- y_j \pm {\rm i})(y_k+
y_j \pm {\rm i}).  
$$
By (\req(etar)), $y_j$'s satisfy the
following system of equations, 
$$
 \frac{- ( 
y_k +
\frac{\rm i}{2})(\Pi(x_k)- y_k{\rm i}
\Phi(x_k)  ) }{( y_k -
\frac{\rm i}{2})( \Pi(x_k) + y_k{\rm i} \Phi(x_k) 
)  } = \prod_{j=1}^n \frac{(y_k- y_j+{\rm i})(y_k+
y_j+{\rm i})}{(y_k- y_j-{\rm i})(y_k+
y_j-{\rm i}) } , \ \ \ 1 \leq k
\leq n ,
$$
or equivalently, 
\be
 \frac{\Pi(x_k)- y_k{\rm i}
\Phi(x_k) }{ \Pi(x_k) + y_k{\rm i}
\Phi(x_k) }= \prod_{j \neq k,  j=1}^n
\frac{(y_k- y_j+{\rm i})(y_k+
y_j+{\rm i})}{(y_k- y_j-{\rm i})(y_k+
y_j-{\rm i}) }, \ \ \ \ 1 \leq k
\leq n .
\ele(Bethe1)

We now consider the equation (\req(SLeqW)) which
arises from (\req(SLe)) by letting  $q \to 1$. 
In the notation of the $q$-gamma function \cite{An:As:Ro}, \cite{Ga:Ra} 
$$
\Gamma_q (y) = \frac{(1-q)^{1-y}(q;
q)_\infty}{(q^y ; q)_\infty }, 
$$
we make the identification 
$$
\frac{1}{(a_je^{i\theta}, a_je^{-i\theta} ;
q)_\infty } =  \frac{\Gamma_q
(-s_j-iy)\Gamma_q
(-s_j+iy)}{(1-q)^{2+2s_j}(q;q)_\infty^2}, \ \ \ a_j = e^{-s_j}.
$$
Note the limiting property $\lim_{q \rightarrow 1}
\Gamma_q (y) = \Gamma (y)$, for a proof see Appendix I in \cite{An}.
The limiting weight function
(\req(qwt)) has the following expression:
\begin{eqnarray}
w(x, \vec{a} ) dx = \frac{
(1-q^{\frac{N}{2}})^2
(q^{\frac{N}{2}};q^{\frac{N}{2}})_\infty^2 
\prod_{l=1}^{2N}\Gamma_q (-s_l-{\rm i}y)\Gamma_q
(-s_l+{\rm i}y) \log q }{(1-q)^{4N+2\sum_js_j}(q;q)_\infty^{4N}
(-{\rm i}Ny)\Gamma_{q^{\frac{N}{2}}}
({\rm i}Ny) } \times \frac{\sin
\theta}{\sin \frac{N \theta}{2}}  d y. \nonumber 
\end{eqnarray}
The fact  $\lim_{\theta \rightarrow 0} \frac{\sin
\theta}{\sin(N\theta/2)}= 2/N$ identifies the following  weight
function as the one corresponding to 
$q=1$, 
\be
w(x, \vec{s} ) = 
\left| \frac{
\prod_{l=1}^{2N}\Gamma
(-s_l+{\rm i}y) }{\Gamma
({\rm i}Ny) } \right|^2.  
\ele(wilsonw)
Consequently  the function $p(x)$ in the
Sturm-Liouville problem (\req(SLeqW)) is given by 
$$
p (x) = w (x, \vec{s}- \frac{\bf 1}{\bf 2}) \qquad \frac{\bf
1}{\bf 2} := (\frac{1}{2}, \ldots, \frac{1}{2}).
$$
The corresponding functions in 
in (\req(wPi)) will be denoted by 
$$
\Pi(x; \vec{a}) = \frac{1}{w(x, \vec{s})}
A w (x, \vec{s}- \frac{\bf 1}{\bf 2}), \
\ 
\Phi(x; \vec{a}) =  \frac{1}{w (x, \vec{s})}
W w (x, \vec{s}- \frac{\bf 1}{\bf 2}).
$$
\begin{thm} \label{thm:XXX} 
For a given $\vec{s}=(s_1, \ldots, s_{2N})$ with
an even $N$, denote
$\varsigma_j$  the
$j$-th elementary symmetric function of $s_i$'s
for $0 \leq j \leq 2N$, $(\varsigma_0 := 1)$. Then 
$ \Pi(x; \vec{a}), x \Phi(x; \vec{a})$ are 
 the polynomials of $x$ of degree at most $N$
with following expressions,
\begin{eqnarray*}
\Pi(x; \vec{a}) &=(-1)^{\frac{N}{2}} \{ 
x^{N}+   
\sum_{j=0}^{N-1} (-1)^j( 
\frac{1}{2} \varsigma_{2j+1}- \varsigma_{2j+2})
x^{N-j-1} \} ,  \\
x \Phi(x; \vec{a}) &=(-1)^{\frac{N}{2}}\{    
\sum_{j=0}^{N-1} (-1)^j( \frac{1}{2}
\varsigma_{2j} - \varsigma_{2j+1})
x^{N-l} + \frac{1}{2} \varsigma_{2N} \}.
\end{eqnarray*}
The roots,  $x_k= y_k^2$, $k=1, \ldots, n$, of a
degree $n$ polynomial solution $f(x)$ of the
Sturm-Liouville problem (\req(SLeqW)) satisfy the
following Bethe Ansatz type relation,
\be
\frac{y_k- \frac{\rm
i}{2}}{y_k+ \frac{\rm i}{2}} 
\prod_{l=1}^{2N}
\frac{y_k+s_l {\rm i}}{ y_k-
s_l {\rm i} } = \prod_{j
\neq k,  j=1}^n
\frac{(y_k- y_j+{\rm i})(y_k+
y_j+{\rm i})}{(y_k- y_j-{\rm i})(y_k+
y_j-{\rm i}) }, \quad 1 \leq k
\leq n.
\ele(gBA1)
\end{thm}
{\it Proof.}
It is easy to see that
\begin{eqnarray} 
\eta_{+} w(x, \vec{s} - \frac{\bf 1}{\bf 2}) 
&=& \frac{
\prod_{j=1}^{2N}\Gamma
(-s_j-{\rm i} y + 1)\Gamma
(-s_j+{\rm i}y) }{\Gamma
(-{\rm i}Ny+\frac{N}{2})\Gamma ({\rm i}Ny
-\frac{N}{2}) }, \nonumber \\
\eta_- w(x, \vec{s} - \frac{\bf 1}{\bf 2}) 
&=& 
\frac{\prod_{j=1}^{2N}\Gamma (-s_j-{\rm i}y)\Gamma
(-s_j+{\rm i}y+1) }{\Gamma
(-{\rm i}Ny-\frac{N}{2}) \Gamma
({\rm i}Ny+\frac{N}{2}) }. \nonumber
\end{eqnarray}
The functional equation of the Gamma function, 
 $\Gamma(z+1) = z \Gamma(z)$, establishes  
the explicit representations for $\Pi$ and $\Phi$, 
\begin{eqnarray}
\Pi(x; \vec{a})  &=&
\frac{(-1)^{\frac{N}{2}}}{2y}\{ (y+\frac{\rm
i}{2})
\prod_{j=1}^{2N}(y -{\rm i}s_j ) +
(y-\frac{\rm i}{2}) 
\prod_{j=1}^{2N} (y +{\rm i}s_j)  \} \nonumber \\
&=& (-1)^{\frac{N}{2}}\{  \sum_{j=0}^{N}
(-1)^j \varsigma_{2j} x^{N-j}+ \frac{1}{2}
\sum_{j=0}^{N-1} (-1)^j  \varsigma_{2j+1}
x^{N-j-1} \}   \nonumber  \\
&=& (-1)^{\frac{N}{2}}\{  x^{N}+   
\sum_{j=0}^{N-1} (-1)^j( 
\frac{1}{2} \varsigma_{2j+1}- \varsigma_{2j+2})
x^{N-j-1} \}   \nonumber \\
\Phi(x; \vec{a}) &=& 
\frac{(-1)^{\frac{N}{2}}}{2y^2 {\rm i}}\{
(y+\frac{i}{2})
\prod_{j=1}^{2N}(y -{\rm i}s_j ) -
(y-\frac{\rm i}{2}) 
\prod_{j=1}^{2N} (y +{\rm i}s_j)  \};  \nonumber \\
&=& (-1)^{\frac{N}{2}} \{ \sum_{j=0}^{N-1}
(-1)^{j+1} \varsigma_{2j+1} x^{N-j-1}+
\frac{1}{2}
\sum_{s=0}^{N} (-1)^j \varsigma_{2j}
x^{N-j-1} \}     \nonumber \\
&=& (-1)^{\frac{N}{2}}\{    
\sum_{j=0}^{N-1} (-1)^j( \frac{1}{2}
\varsigma_{2j} 
- \varsigma_{2j+1}) x^{N-j-1} + \frac{1}{2}
\varsigma_{2N}x^{-1}\}. \nonumber
\end{eqnarray}
Hence we obtain the expressions of
$\Pi(x; \vec{a}), x \Phi(x; \vec{a})$. By the
 first expressions for $\Pi(x; \vec{a}), 
\Phi(x; \vec{a})$ in terms of $y$ in the above right hand sides, one
can easily derive the following identities:
$$
\Pi(x; \vec{a})  - \Phi(x; \vec{a}) y{\rm i}  
= 
\frac{(-1)^{\frac{N}{2}}}{y}
(y-\frac{\rm i}{2}) 
\prod_{j=1}^{2N} (y + {\rm i}s_j) , \ \ \
\Pi(x; \vec{a})  + \Phi(x; \vec{a}) y{\rm i} 
= 
\frac{(-1)^{\frac{N}{2}}}{y} (y+\frac{\rm i}{2})
\prod_{j=1}^{2N}(y -{\rm i}s_j ), 
$$
hence
$$
\frac{\Pi(x; \vec{a}) - \Phi(x; \vec{a}) y{\rm i} }{
\Pi(x; \vec{a}) + \Phi(x; \vec{a}) y{\rm i} }
= 
\frac{y_k- \frac{\rm
i}{2}}{y_k+ \frac{\rm i}{2}} \prod_{j=1}^{2N}
\frac{  y +s_j{\rm i} }{ y -s_j{\rm i } }.
$$
By the above relation and
(\req(Bethe1)), we obtain (\req(gBA1)). 
$\Box$ \par \vspace{.2in} 

\noindent
{\bf Remark.} For the Bethe Ansatz equations (\req(gBA1))
with $N$ odd, one can reduce the problem to the
 above theorem for some even $N'$ by
adding certain zero-value $s_j$'s.
By the similar method, one enables to apply the
above theorem to  Bethe Ansatz problem  of the 
following type with $s_1, \ldots, s_M \in \CZ$
and $ M \in \NZ$,
\be
\prod_{l=1}^{M}
\frac{y_k+s_l {\rm i}}{ y_k-
s_l {\rm i} } = \prod_{j
\neq k,  j=1}^n
\frac{(y_k- y_j+{\rm i})(y_k+
y_j+{\rm i})}{(y_k- y_j-{\rm i})(y_k+
y_j-{\rm i}) }, \quad 1 \leq k
\leq n,
\ele(gBA2)
$\Box$ \par \vspace{.2in} 

For a positive half-integer $s$, it is known that
the spin
$s$ XXX model of an even size $L$ with the
periodic boundary condition has the following
Bethe Ansatz equations:
\be
(\frac{\lambda_k + s {\rm i}}{\lambda_k
- s {\rm i}})^L = \prod_{j=1, j \neq k}^l
\frac{\lambda_k - \lambda_j + {\rm i}}{\lambda_k
- \lambda_j - {\rm i}}, \ \ \lambda_k \in \CZ, \ \ \ k = 1 , \ldots, l.
\ele(BXXX)
In the antiferromagenetic case, the ground state
is on the sector $l = \frac{L}{2}$; conjecturally
there is the unique real
solution of the Bethe Ansatz equations. Note that $\{ -
\lambda_j \}_{j=1}^l$ is a solution of
(\req(BXXX)) whenever $\{ 
\lambda_j \}_{j=1}^l$ is a solution. Hence for the
ground state, the roots
$\lambda_j$s are expected to be real and invariant
under the sign-change ( up to permutation
of the indices
$j$). Having this ground state conjecture in mind,
we now consider a general problem of (\req(BXXX))
for $l
\equiv
\frac{L}{2} \pmod {2} $ with  roots 
invariant under the change of sign,
$\lambda_j \mapsto -\lambda_j$, i.e., 
$\lambda_j$s with the following form:
\begin{eqnarray}
&L = 4M+2 , & \{ \lambda_j \}_{j=1}^{l} = \{0,
\pm y_1, \ldots , \pm y_n \} ; \label{XXXO} \\
&L = 4M , & \{ \lambda_j \}_{j=1}^{l} = \{
\pm y_1, \ldots , \pm y_n \} \ \label{XXXE} .
\end{eqnarray}
In this situation, one can link the Bethe Ansatz equations 
(\req(BXXX)) of XXX model to
the Sturm-Liouville problem previously discussed in this
section.   In the case (\ref{XXXO}), the relation
(\req(BXXX)) becomes
$$
\frac{(y_k -
\frac{\rm i}{2})(y_k-{\rm i})(y_k + s {\rm
i})^{4M+2}}{(y_k + \frac {\rm i}{2})(y_k+{\rm
i})(y_k-s{\rm i})^{4M+2}} =
\prod_{j=1, j \neq k}^n
\frac{(y_j - y_k + {\rm i})(y_j + y_k + {\rm
i})}{( y_j - y_k - {\rm i})( y_j - y_k + {\rm i})
}, \ \
y_k \in \CZ \setminus \{ 0 \}, \ \ \ k = 1,
\ldots, n, 
$$
in which case, we set the $N$, $s_j$'s in Theorem
\ref{thm:XXX} as follows:
\bea(llll)
L = 4M+2 , & s \neq 1: & N=2M+2    , & s_1=0, \ \
s_2=-1, s_3=
\cdots= s_{2N} = s , \\
&&&\varsigma_{2N}=0, \ 
\varsigma_j = {2N-2
\choose j}s^j- { 2N-2 \choose j-1}s^{j-1},
\ \ 
 0 \leq j \leq 2N-1 ; \ 
\\ & s=1: & N=2M+1  , &
s_1 = 0 , \ \ s_2 =
\dots = s_{2N}= 1 , \\
&&& \varsigma_{2N}=0, \ \ \varsigma_j = {2N-1
\choose j } \ \ 
0 \leq j \leq 2N-1.
\elea(sO)
In the case (\ref{XXXE}), one has  
\be
\frac{(y_k - \frac{\rm
i}{2})(y_k+s{\rm i})^{4M}}{(y_k + \frac {\rm
i}{2})(y_k-s{\rm i})^{4M}} =
\prod_{j=1, l \neq k}^n
\frac{(y_j - y_k + {\rm i})(y_j + y_k + {\rm
i})}{( y_j - y_k - {\rm i})( y_j - y_k + {\rm i})
}, \ \
y_k \in \CZ \setminus \{ 0 \}, \ \ \ k = 1 ,
\ldots, n,
\ele(XXXBs)
in which case,  $N, s_j$ in Theorem
\ref{thm:XXX} are given by
\be
L = 4M, \ \  N=2M, \quad  s_1= \cdots = s_{2N}
= s
\ , \quad \varsigma_j = {2N \choose j }s^j \quad {\rm for}
 \ 0 \leq j \leq 2N.
\ele(sE)
For both situations, the relations are of the form (\req(gBA2))
for suitable $s_l$'s. Note that the non-zero
condition of $y_k$s in (\ref{XXXO}), (\ref{XXXE}) for  the 
corresponding
polynomial solution $f(x)$ of (\req(SLeqW))
requires one further constraint, namely $f(0)
\neq 0$.  The ground state of antiferromagenetic
spin
$s$ XXX model of size $L$ is governed by the
real root solution of the above equations for
$n=\frac{L}{2}$.

We now consider the case $s=\frac{1}{2}$ and
discuss some mathematical problems relevant to the 
physics of the system. The Hamiltonian is given
by $H_{\rm XXX}$ in (\req(XXX)). 
We will discuss the Bethe Ansatz equations 
(\req(XXXBs)) with
$s=\frac{1}{2}$ in (\req(sO)) (\req(sE)). To
illustrate the mathematical content connecting to
the Bethe Ansatz equations for the 
Hamiltonian $H_{\rm XXX}$, we consider the case
$L=2, 4$. For $L=2$ in 
(\ref{XXXO}), by 
$$
 \left|\Gamma({\rm i}y)\right|^2 =
\frac{\pi}{y \sinh \pi  y},\ \ \ 
\left| \Gamma(\frac{-1}{2}+ {\rm i}y)\right|^2=
\frac{4 \pi}{(4y^2+1)\cosh \pi y},
$$
the weight function is expressed by 
$$
w(x) = \left| \frac{ \Gamma ({\rm
i}y)\Gamma(1+ {\rm i}y ) \Gamma(\frac{-1}{2}+{\rm
i}y)^2 }{\Gamma (2{\rm i}y)}\right|^2 =\frac{(2\pi)^3 y}
{(x+\frac{1}{4})^2 \sinh 2 \pi y}.
$$
By (\req(sO)), the corresponding  Sturm-Liouville
problem (\req(SLeqW)) is governed by the equation
$$
(8x^2 + 6 x + 1) W^2 f (x) +
(4x + 1) AWf (x) = \lambda
f(x)
$$
with $\lambda = 4n(2n-1)$ except $n = 1$,
in which case $\lambda= 4$.  
For $L=4$ in (\ref{XXXO}), by (\req(sE)) we have
the following weight function and the
Sturm-Liouville problem,
\begin{eqnarray*}
w(x) = \left|\frac{ \Gamma(\frac{-1}{2}+{\rm
i}y)^4 }{\Gamma (2{\rm i}y)}\right|^2 = \frac{4 \pi^3 y 
\sinh  \pi y}{(x+\frac{1}{4})^4
\cosh^3 
\pi y}  \qquad \quad   \\
-x(32x^2 - 16 x -6) W^2 f
(x) + (48 x^2 +
8x-1) AWf (x) =
(\lambda x + \mu) f(x)
\ 
\end{eqnarray*}
with $\lambda  = -16n (2n-1)$ except $n = 1$,
in which case
$\lambda = 48$. For $n=1$, we have $\mu= 4$
and $f(x)= x + \frac{1}{4}$.  Note
that due to the  parameters in the above weight
functions,  the Sturm-Liouville problems we
encounter here does not make 
the Wilson polynomials orthogonal, as in 
\cite{As:Wi}. However, our empirical computation
for the case $L=2$ has indicated that the $f(x)$'s are
real polynomials with $f(0) \neq 0$. 
For the general size $L$, the
weight function of the Sturm-Liouville problem
connecting to XXX model in the context of 
discussions in this section has the following expressions:
\begin{eqnarray}  
L=4M+2,  &{}&  w(x) = \left| \frac{ \Gamma ({\rm
i}y)\Gamma(1+ {\rm i}y ) \Gamma(\frac{-1}{2}+{\rm
i}y)^L }{\Gamma ((\frac{L}{2}+1){\rm i}y)}\right|^2 =
\frac{(\frac{L}{2}+1)\pi^{L+1}y \sinh
(\frac{L}{2}+1) \pi y }{
(x+\frac{1}{4})^L \sinh^2
\pi  y \cosh^L \pi y},  \nonumber \\
L=4M , &{}&  w(x) = \left|\frac{ \Gamma(\frac{-1}{2}+{\rm
i}y)^L }{\Gamma (\frac{L}{2}{\rm i}y)}\right|^2 \qquad  \qquad
\qquad =
\frac{
\frac{L}{2}\pi^{L-1}y\sinh \frac{L}{2}\pi 
y}{(x+\frac{1}{4})^L\cosh^L \pi y}. \nonumber 
\end{eqnarray}
Nevertheless a sound  mathematical treatment
of their corresponding Sturm-Liouville equations
appears still lacking now, and it remains a
difficult task to obtain substantial knowledge
incorporating the physical applications associated to XXX
model. However The quest of a such mathematical
theory would be a necessary one in order to
understand the  essential features of the Bethe
Ansatz equations for the Heisenberg XXX spin chain.

\section{Conclusions and Perspectives}
In this paper we have 
provided a one to one correspondence between polynomial solutions to 
Sturm-Liouville
type equation involving the\aw operator and solutions of the algebraic Bethe
Ansatz equations of XXZ and XXX models.  We have also started 
a preliminary  study of the 
mathematical problems arising  from the  physics of the Bethe Ansatz. 
In doing so we have reduced the nonlinear problem of solving the Bethe 
Ansatz equations to the linear problem of finding polynomial solutions 
to certain linear second order equations in terms of the\aw or Wilson operator. 
Thus the physics of XXZ model has indeed raised some mathematical 
questions which demand the need for a systematic
 mathematical development  in the theory of the corresponding $q$-difference
equations. The mathematical solution of those questions will  
lead to a better  understanding of some interesting physical quantities of XXZ and XXX models.   
 We intend to continue this study in future and partial results 
 obtained so far are indeed very promising.

\bigskip

{\bf Acknowledgments} M. Ismail acknowledges 
financial support from Liu Bie Ju Center of Mathematical Sciences in Hong Kong, and 
the Academia Sinica of Taiwan, where this work started. This paper was partially supported by 
NSF grant DMS 99-70865 and NSC grant 89-2115M001037 of Taiwan.

\end{document}